\documentclass[aps,showpacs,superscriptaddress,12pt,tightenlines]{revtex4}
\usepackage{times}
\usepackage{amsmath}
\usepackage{graphicx}
\usepackage{dcolumn}
\usepackage{bm}
\usepackage{color}
\usepackage{subfigure}
\usepackage{hyperref}
\newcommand{\BR}{{\cal B}}
\newcommand{\ee}{e^+e^-}
\newcommand{\y}{\psi(4230)}

\newcommand{\EE}{e^+e^-}

\newcommand{\psp}{\psi(3686)}
\newcommand{\psip}{\psi(3686)}
\newcommand{\psipp}{\psi(3770)}
\newcommand{\jpsi}{J/\psi}
\newcommand{\piz}{\pi^0}
\newcommand{\ddb}{D\bar{D}}
\newcommand{\zc}{Z_c(3900)}

\newcommand{\ppb}{p\bar{p}}
\newcommand{\ppbpiz}{\ppb\piz}

\newcommand{\degree}{^{\circ}}
\newcommand{\panda}{\overline{\text{P}}\text{ANDA}}
\newcommand{\GeV}{\text{GeV}}

\begin{document}

\title{\boldmath
Improved description of $\EE\to p\bar{p}\pi^0$ cross section line shape and 
more stringent constraints on $\psipp$ and $\y\to \ppbpiz$}
\author{Ya-Qian Wang}
 \email{whyaqm@hbu.edu.cn}
 \affiliation{Department of Physics, Hebei University, Baoding 071002, China}
 \affiliation{Institute of High Energy Physics, Chinese Academy of Sciences,
 Beijing 100049, China}
\author{Chang-Zheng Yuan}
 \email{yuancz@ihep.ac.cn}
 \affiliation{Institute of High Energy Physics, Chinese Academy of Sciences,
 Beijing 100049, China}
 \affiliation{University of Chinese Academy of Sciences, Beijing 100049, China}

\date{\today}

\begin{abstract}

By using the cross sections of $\EE\to \ppbpiz$ measured at the 
center-of-mass energies from 3.65 to 4.60~GeV, we find that 
the line shape well agrees with pure continuum production 
parametrized by a power law function and the significance of 
both $\psipp$ and $\y\to \ppbpiz$ is less than $0.4\sigma$. 
We set more stringent constraints on the charmless decays
of the charmonium state, $\psipp$, and the charmoniumlike state, $\y$,
to $\ppbpiz$ than in previous measurements. The data are also used
to estimate the cross sections of $\ppb\to \piz\psipp$ and
$\ppb\to \piz\y$ that are essential for planning the data taking
of the $\overline{\text{P}}\text{ANDA}$ experiment.

\end{abstract}

\pacs{14.20.Dh, 13.25.Gv, 13.66.Bc}

\maketitle

The charmonium state $\psipp$, lying about 40~MeV above the open charm threshold, 
is expected to decay dominantly into the Okubo-Zweig-Iizuka allowed final state, $\ddb$, together
with a small fraction of hadronic and radiative transitions into final states 
with a lower mass charmonium state. However, the total cross section 
$\sigma(\ee\to\psipp)$ measured by counting the total number of inclusive 
hadronic events is not saturated by $\ddb$ measured by counting the number
of charged and neutral $\ddb$ pairs~\cite{c_nonDD_bes,c_nonDD_cleo}. 
The hadronic transitions and the radiative transitions of the $\psipp$
can be calculated reliably in the quark models~\cite{quarkmodels} 
and have been measured precisely in experiments~\cite{c_pdg_2020}.
This indicates that noticeable charmless decays of the $\psipp$, i.e., 
$\psipp\to \text{\it light hadrons}$, may exist. 
On the other hand, people did elaborate studies to search for exclusive 
charmless decays and failed to find any mode with a large branching
fraction~\cite{c_nonDD_ex_BES,c_nonDD_ex_CLEO}. 

The large rate of charmless decays of the $\psipp$ can also be accommodated
in theoretical models. Among many theoretical efforts trying to solve the 
``$\rho\pi$ puzzle'' in $\jpsi$ and $\psp$ decays observed by the 
Mark-II experiment~\cite{rhopi_markII}, the 2S-1D charmonium mixing 
scenario~\cite{c_JLRosener} relates a partial width in $\psipp$ decays 
into any final state to the corresponding partial widths in $\jpsi$ 
and $\psip$ decays~\cite{c_Rosener_WYM}, as both $\psip$ and $\psipp$
are the mixtures of the 2S and 1D charmonium states. As a result, 
a large $\psipp\to \text{\it light hadrons}$ is allowed in this model,
and the measurement of the $\psipp$ decays can test the 
model predictions. The process $\psipp\to\ppb\piz$ is one of the possible 
exclusive channels contributing to the charmless decays of the $\psipp$. 

The observation of the charmoniumlike states such as the $X(3872)$~\cite{X3872}, 
the $Y(4260)$~\cite{Y4260}, and the $\zc$~\cite{Zc3900} indicates 
that the hadrons are more complicated 
than the expectations in quark models. The exotic properties of these
states such as very close to open charm thresholds and large coupling to
hidden-charm final states may suggest they are hadronic states beyond 
the conventional quark model~\cite{reviews}. 
Search for their decays into light hadrons
will also shed light on their nature. Considering the small coupling 
to open charm final states, we may even expect a state like the $\y$~\cite{Y4230}
(the dominant component of the $Y(4260)$ structure) has a larger 
decay rate to the charmless final state than a charmonium state does.
These discussions can be extended to other vector charmoniumlike
states such as the $Y(4360)$ and $Y(4660)$~\cite{Y4660}.

If indeed $\psipp\to \ppbpiz$ and/or $\y\to \ppbpiz$ are observed,
one may calculate $\sigma(\ppb\to\piz\psipp)$ and $\sigma(\ppb\to\piz\y)$
by using the cross symmetry as has been done in Ref.~\cite{PhysRevD.73.096003}. 
These cross sections serve as an essential input for the $\panda$ 
(Anti$\underline{\text{P}}$roton $\underline{\text{An}}$nihilations experiment
at $\underline{\text{Da}}$rmstadt), which plans to study charmonium
and charmoniumlike states produced in $\ppb$ annihilation.

To study the resonance contribution in the $\ee\to\ppb\piz$ channel, 
the process $\psipp\to\ppb\piz$ can hardly be determined
without a good knowledge of the continuum production~\cite{guoyp}. 
The continuum cross sections of $\ee\to\ppb\piz$ are calculated 
in the energy range from threshold up to $\sqrt{s}=4.2$~GeV, 
applying the conservation of the hadron electromagnetic currents 
and the P-invariance of the hadron electromagnetic interaction~\cite{c_ppbpiz_th}. 
We focus on the study of $\ee\to\ppb\piz$ in this paper, 
and aim to extract both the resonance and the continuum contributions.

The process of $\EE\to \ppbpiz$ is studied in the vicinity of
the $\psipp$ by the BESIII experiment in 2014~\cite{BESIII-2014} 
(labeled with ``2014'' hereinafter).
Two indistinguishable 
solutions~\cite{c_zhuk} for the cross section of $\psipp\to \ppbpiz$ 
are extracted, and the maximum cross section of $\ppb \to \psipp \pi^0$ is expected at a 
center-of-mass energy (CME) of 5.26~GeV  
using  a constant decay amplitude 
approximation~\cite{PhysRevD.73.096003}.
The same process is studied in the energy range $\sqrt{s} = 4.008\sim 4.600$~GeV 
with the BESIII data taken at 13 CMEs~\cite{BESIII-2017} 
(labeled with ``2017'' in the context), and 
no significant resonance is observed.

The above two analyses study the $\EE\to \ppb\piz$ process 
from $\sqrt{s}=3.65\sim 4.60$~GeV, and in principle the data can be combined
to extract a better estimation of the physics quantities. 
By combining the measurements, we benefit not only from a better 
description of the continuum contribution but also from improvement 
on the fit result in the vicinity of the $\psipp$ and the $\y$.
In a word, besides feeding the $\panda$ experiment by estimating the cross
sections $\sigma(\ppb\to\piz\psi)$, the results could also 
provide further insights into the puzzling question on the mechanisms of
non-$\ddb$ transitions for $\psipp$ and shed light on the understanding of the
$\y$ and other vector charmoniumlike states.

Both the BESIII analyses fit the cross sections with the formula
\begin{eqnarray}
  \sigma(m)= & \left| \sqrt{\sigma_{\rm con}} + \sqrt{f_{\sigma_{\psi}}} \frac{M\Gamma}{m^2-M^2+iM\Gamma} e^{i\phi} \right|^2, 
\label{eq:cross_section}
\end{eqnarray} 
where $\sigma_{\rm con}=C/s^{\lambda}$ represents the continuum amplitude with the 
unknown exponent $\lambda$ and a constant $C$; 
the resonance $\psi$ has a fixed mass $M$ and total width $\Gamma$~\cite{c_pdg_2020}; 
the factor $f_{\sigma_{\psi}}=\sigma(\EE\to \psi\to \ppbpiz)$ is the peak cross section of the resonance $\psi$; 
the parameter $\phi$ describes the relative phase between the continuum 
and resonance amplitudes. 

\begin{figure*}[htbp]
  \begin{center}
    \includegraphics[width=\textwidth]{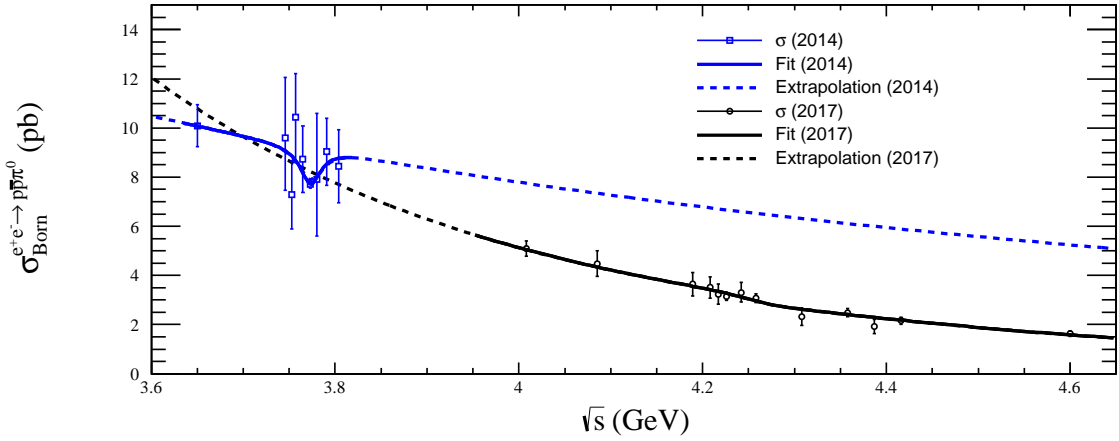}
    \caption{\label{f_fit0} Extrapolation of the BESIII fit results, 
    the 2014~\cite{BESIII-2014} and 2017~\cite{BESIII-2017} analyses 
    give very different estimation of the continuum amplitude. } 
  \end{center}
\end{figure*}

The two analyses study the same final state but at different energy ranges.
The cross sections are fitted with the same model but with different
resonances. The parameters ($C$, $\lambda$) describing the continuum 
contribution are expected to be the same.
However, an obvious discrepancy exists between the two best fits in
the regions dominated by the continuum amplitude, as shown in Fig.~\ref{f_fit0}
and indicated by the very different common parameters listed in Table~\ref{t_compar}.

\begin{table*}[htbp]
  \caption{\label{t_compar} Fit parameters in the two BESIII analyses~\cite{BESIII-2014,BESIII-2017}. 
  The resonance parameters of the $\psipp$ and $\y$ are fixed in the fits.}
  \begin{tabular}{c|c|c|c|c|c}
    \hline\hline
    Data& $\sqrt{s}$ (GeV)& $C$ ($\text{ GeV}^{2\lambda}\text{pb}$)&  $\lambda$&  $M_\psi$ (GeV)& $ \Gamma_\psi$ (GeV) \\
    \hline
    2014& $3.650-3.804$&  $(0.4\pm0.6)\times 10^3$&  $1.4\pm0.6$& $3.77315$&  $0.0272$\\
    2017& $4.008-4.600$&  $(5.4\pm5.3)\times 10^5$&  $4.2\pm0.4$& $4.251$&  $0.12$\\
    \hline\hline
  \end{tabular}
\end{table*}

We try to do a combined fit to the measurements presented in the 
two analyses~\cite{BESIII-2014,BESIII-2017}. Table~\ref{t_data} 
shows a compilation of the cross sections.
We do a least $\chi^2$ fit with
$$
\chi^2 = \sum_{i=1}^{N}
         \frac{(\sigma^{\rm meas}_i-\sigma^{\rm fit}(m_i))^2}
              {(\Delta\sigma^{\rm meas}_i)^2},
$$
where $\sigma^{\rm meas}_i\pm \Delta\sigma^{\rm meas}_i$ is the
dressed cross section from experimental measurement, 
and $\sigma^{\rm fit}(m_i)$ is the cross
section value calculated from the model below with the parameters
obtained from the fit. Here $m_i$ is the CME that corresponds to the $i$th
one of all the $N$ energy points. 
We only use the statistical errors in our fits since the systematic
errors ($\sim 6.5\%$) for all the data points are correlated. 

\begin{table}[htbp]
\centering
  \caption{\label{t_data}$\EE\to\ppbpiz$ Born cross sections measured by 
  the BESIII experiment~\cite{BESIII-2014,BESIII-2017}. The first errors 
  are statistical and second ones systematic. $\frac{1}{|1-\Pi|^2}$ is the vacuum 
  polarization factor to obtain the dressed cross sections~\cite{c_vp_cmd}.}
\begin{tabular}{c c c}
\hline \hline 
  ~~~$\sqrt{s}$~(GeV)~~~ &  $\sigma(\EE\to \ppbpiz)$~(pb)&  ~~~$\frac{1}{|1-\Pi|^2}$~~~ \\
\hline
  3.650 &   $10.09 \pm 0.84 \pm 0.16$&  1.0196  \\ 
  3.746 &   $ 9.60^{+2.45}_{-2.12}  \pm 0.16 $& 1.0596 \\ 
  3.753 &   $7.28 \pm 1.38 \pm 0.12$& 1.0573\\ 
  3.757 &   $ 10.44 \pm 1.77\pm 0.17 $& 1.0564 \\ 
  3.765 &   $ 8.73 \pm 1.35 \pm 0.14 $& 1.0558\\ 
  3.773 &   $ 7.71 \pm 0.09 \pm 0.13$&  1.0598 \\ 
  3.780 &   $7.92^{+2.66}_{-2.31} \pm 0.13$&  1.0611 \\ 
  3.791 &   $9.03 \pm 1.35 \pm 0.15 $&  1.0592 \\  
  3.804 &   $ 8.44\pm 1.48 \pm 0.14 $&  1.0573 \\ 
\hline
  4.008  &$5.09\pm0.18_{-0.24}^{+0.26}$&  1.004\\
  4.085  &$4.47\pm0.46_{-0.21}^{+0.27}$&  1.052\\
  4.189  &$3.64\pm0.43_{-0.19}^{+0.18}$&  1.056\\
  4.208  &$3.52\pm0.39_{-0.22}^{+0.17}$&  1.057\\
  4.217  &$3.24\pm0.37\pm0.18$& 1.057\\
  4.226  &$3.15\pm0.08\pm0.14$& 1.056\\
  4.242  &$3.30\pm0.36_{-0.15}^{+0.19}$&  1.056\\
  4.258  &$3.08\pm0.10_{-0.15}^{+0.14}$&  1.054\\
  4.308  &$2.32\pm0.33_{-0.10}^{+0.15}$&  1.053\\
  4.358  &$2.48\pm0.11_{-0.12}^{+0.13}$&  1.051\\
  4.387  &$1.92\pm0.26\pm0.10$& 1.051\\
  4.416  &$2.16\pm0.10_{-0.11}^{+0.10}$&  1.053\\
  4.600  &$1.63\pm0.08\pm0.08$& 1.055\\
  \hline \hline
\end{tabular}
\label{tab:results_overview}
\end{table} 

\begin{figure*}[htbp]
  \begin{center}
    \includegraphics[width=\textwidth]{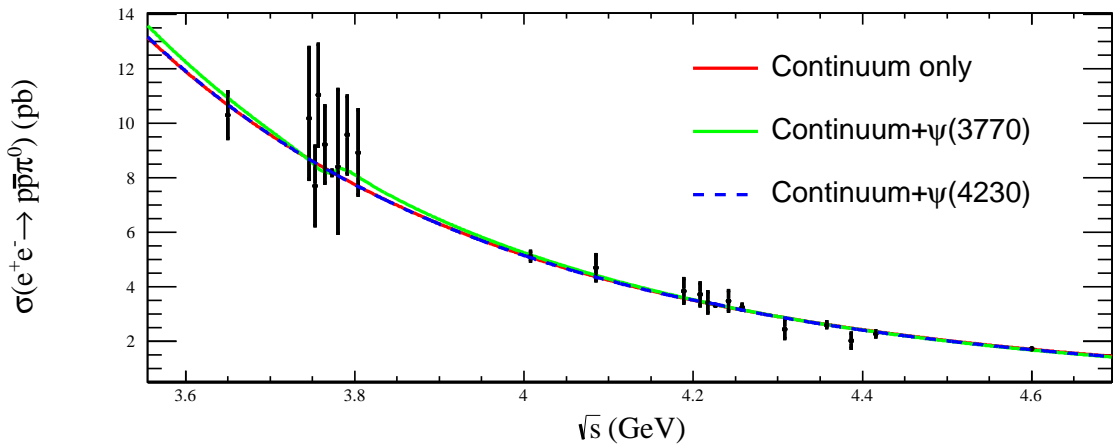}
  \end{center}
  \caption{\label{f_com_stat}Combined fits to the $\EE\to\ppbpiz$ cross 
  sections~\cite{BESIII-2014,BESIII-2017}. The red curve is the fit 
  with continuum amplitude only, the green curve is the fit with coherent
  sum of the continuum and the $\psipp$ amplitudes, and the dashed blue
  curve is the fit with coherent sum of the continuum and the $\y$ amplitudes.}
\end{figure*}

First, we fit the cross sections from 3.65 to 4.60~GeV with 
the continuum amplitude only. The fitted result is shown as the 
red curve in Fig.~\ref{f_com_stat}, and the fitted parameters are listed in Table~\ref{t_com_fit}.
The goodness-of-fit is $\chi^2/$NDF=13.0/20 (NDF is the number of degrees of freedom), 
corresponding to a confidence level of 88\%, a very good fit. 

We then fit the cross sections considering possible resonance contributions.
The continuum amplitude and the resonance ($\psipp$ or $\y$) amplitudes are 
added coherently, 
$$
\sigma(m)=  \left| \sqrt{\sigma_{\rm con}} + BW(m) e^{i\phi} \right|^2,
$$
where 
$BW(m)=\frac{\sqrt{12\pi\Gamma_{\EE}\Gamma_{\rm tot}\times\BR(\psi\to\ppbpiz)}}
{m^2-M^2+i M\Gamma_{\rm tot}}$ is a Breit-Wigner function to describe the
resonance amplitude with a mass $M$,
total width $\Gamma_{\rm tot}$, electronic partial width $\Gamma_{\EE}$, and
the branching fraction $\BR(\psi\to\ppbpiz)$.
The continuum term $\sigma_{\rm con}$ is defined the same as before. 
In the fits, both the mass and the total width 
of the resonances are fixed according to the PDG~\cite{c_pdg_2020},
and the product $\Gamma_{\EE}\times\BR(\psi\to\ppbpiz)$ is a free parameter.
The continuum term $\sigma_{\rm con}$ and the relative phase between 
the continuum and the resonance amplitudes, $\phi$, are also float parameters.

The $\psipp$ or the $\y$ resonance amplitude is added to the fit. 
The fit results are shown as the solid green and dashed blue curves in 
Fig.~\ref{f_com_stat} and listed in Table~\ref{t_com_fit}. 
The goodness-of-fit, $\chi^2$/NDF, is 
$12.3/18$ and $12.8/18$ for $\psipp$ and $\y$, corresponding to 
a confidence level of 83\% and 80\%, respectively.
The statistical significance of the $\psipp$ is $0.36\sigma$ and
that for the $\y$ is $0.11\sigma$, by comparing the differences
in the $\chi^2$ and in the NDF.
Two solutions are found in each fit, one of which
has a well determined magnitude and interferes with the
continuum amplitude destructively ($\phi\sim 270\degree$), 
and the other solution comes with large uncertainty and agrees
with zero with statistical uncertainties. 
The two solutions of $\Gamma_{\EE}\times\BR$ may differ
by several orders of magnitude for both $\psipp$ and $\y$.

\begin{table*}[htbp]
  \caption{\label{t_com_fit}Combined fit results for the three configurations. 
  The subscripts of $\psipp$ and $\y$ indicate the two solutions of the fits.
  The errors are statistical only. There is an additional 6.5\% systematic error 
  in $\Gamma_{\EE}\times\BR$ and $C$ due to the common systematic error in
  the cross section measurements. } 
  \begin{tabular}{c|c|c|c|c|c}
    \hline\hline
    Fit&  $\chi^2$/NDF& $\Gamma_{\EE}\times\BR$ (eV)& $\phi$ ($\degree$)&  $C$ ($10^5\GeV^{2\lambda}$ pb)&  $\lambda$\\
    \hline
    Continuum&  13.0/20&$\ldots$&$\dots$& $3.07\pm0.58$&  $3.96\pm0.07$\\
    \hline
    $\psipp_1$& 12.3/18& $(0.28\pm1.40)\times 10^{-3}$&  $340\pm54$&  $4.1\pm1.8$&  $4.07\pm0.15$\\
    $\psipp_2$& 12.3/18& $0.875\pm0.017$&  $270.1\pm2.4$& $4.1\pm1.8$&  $4.07\pm0.15$\\
    \hline
    $\y_1$& 12.8/18& $(0.88\pm4.20)\times 10^{-5}$&  $78\pm202$& $3.22\pm0.83$&  $3.98\pm0.10$\\
    $\y_2$& 12.8/18& $0.820\pm0.016$&  $268.7\pm1.1$&  $3.23\pm0.84$&  $3.98\pm0.10$\\
    \hline\hline
 \end{tabular}
\end{table*}

By combining the two BESIII measurements, the fit parameters, especially 
those describing the continuum are further constrained. 
Our first fit without any resonance indicates in another way 
the extremely small significance of the $\psipp$ or $\y$. 
We notice that the parameters $C$ and $\lambda$ are relatively stable 
in all the fits within uncertainties. 
The theoretical predictions on the continuum amplitude vary significantly 
in both the shape and the magnitude with different parametrizations 
(``new'' and ``old'')~\cite{c_ppbpiz_th}. The new one favored by the BESIII data
follows a power law with $C=2.1\times 10^3~\GeV^{2\lambda}$ and $\lambda=3.58$ in the CME 
from 3.6 to 4.2~$\GeV$~\cite{c_ppbpiz_th}. However, the absolute cross sections, 
reflected by $C$, is still below the BESIII data by two orders of magnitude. 
The $\lambda$, describing the slope of the line shape, is sensitive to the choice of 
parametrization and can be further constrained by our combined fit.

Although not significant, the central value of the branching fraction of 
$\psipp\to\ppbpiz$ is also extracted to be either 
$(1.1\pm5.4\pm0.1\pm0.1^{\Gamma_{\EE}})\times 10^{-6}$ or 
$(0.33\pm 0.01\pm0.03\pm0.03^{\Gamma_{\EE}})$\% using 
$\Gamma_{\EE}=(0.262\pm0.018)$~keV from the PDG~\cite{c_pdg_2020},
where the errors are statistical, systematic, and from the uncertainty 
of quoted $\Gamma_{\EE}$~\cite{c_pdg_2020}.
Very recently, authors of Ref.~\cite{Achasov:2021adv} obtained 
$\Gamma_{\EE}=(0.19\pm0.04)$~keV for the $\psipp$ by taking into account 
the contributions of the mixed $\psipp$ and $\psip$ resonances. 
With this $\Gamma_{\EE}$, the corresponding branching fraction is 
calculated as $(1.5\pm7.4\pm0.1\pm0.4^{\Gamma_{\EE}})\times 10^{-6}$
or
$(0.46\pm0.01\pm0.03\pm0.10^{\Gamma_{\EE}})$\%.  
All the results are within their large uncertainties either from our fit 
or from the $\Gamma_{\EE}$. 
Obviously, the branching fraction of $\psipp\to\ppbpiz$ highly depends on 
the choice of the $\Gamma_{\EE}$ in our calculation, and any reliable 
$\Gamma_{\EE}$ can be taken as input to extract the branching fraction. 
The two solutions suggest 
either noticeable or negligible charmless decays of $\psipp\to \ppbpiz$. 
As one of the charmless decays of the $\psipp$, the branching fraction can 
also be estimated considering the mixing with the $\psip$ 
due to tensor forces and coupling to charmed meson pairs~\cite{c_JLRosener}. 
This S- and D-wave charmonium mixing model accounts for the ``$\rho\pi$ puzzle'' 
by converting expected $\psip\to \rho\pi$ to corresponding $\psipp$ partial widths.
Taking the $\jpsi$ and the $\psip$ decays into the same final state as input, 
the branching fraction of $\psipp\to\ppbpiz$ is predicted in the range 
[$(2.9\pm 4.9)\times 10^{-7}$, $(15.3\pm 1.5)\times 10^{-5}$] assuming the mixing angle
$\theta=(12\pm 2)\degree$. One of our fit results falls
in the range of this prediction, while the other solution is beyond 
the range by one order of magnitude.
If we use another mixing angle $\theta=-(27\pm 2)\degree$, 
the model predicts the branching fraction to be in the range
[$(2.4\pm 1.9)\times 10^{-7}$, $(2.9\pm 0.3)\times 10^{-5}$]. 
Compared to the results with $\theta=12\degree$, the range 
becomes narrower, and the maximum decay rate of $\psipp\to\ppb\piz$ 
is much smaller.

The $\panda$ experiment produces neutral state with any conventional quantum numbers 
($J^{PC}$) through $\ppb$ reactions. While exotic states is also available in association
with an additional meson, e.g., $\ppb\to\piz X$, where $X$ is an exotic hybrid or a 
charmonium~\cite{PhysRevD.73.096003,Pire:2013jva}. The branching fraction of $\psipp\to \ppb\piz$ 
can be taken as input for the constant decay amplitude model~\cite{PhysRevD.73.096003}
to calculate the cross section of $\ppb\to \piz\psipp$.
The maximum cross sections predicted by the model is 
$\sigma(\ppb\to\piz\psipp)=(0.04\pm 0.20\pm0.01\pm0.01^{\Gamma_{\EE}})$~nb 
or $(118\pm 6\pm12\pm12^{\Gamma_{\EE}})$~nb 
at $\sqrt{s}=5.26$~GeV, both agree with previous results~\cite{BESIII-2014}. The cross sections are 
sensitive to the $\Gamma_{\EE}$ in the same way as the
branching fraction of $\psipp\to\ppbpiz$.
If we use $\Gamma_{\EE}=(0.19\pm0.04)$~keV~\cite{Achasov:2021adv}, 
the above cross section becomes
$(0.05\pm0.27\pm0.01\pm0.02^{\Gamma_{\EE}})$~nb 
or $(164\pm 7\pm13\pm37^{\Gamma_{\EE}})$~nb, which are larger by 25\% and 39\% 
compared with the results with $\Gamma_{\EE}$ from the PDG~\cite{c_pdg_2020}. 
The production rate of $\ppb\to\piz\y$ reaches the peak at $\sqrt{s}=5.9$~GeV,
which exceeds the energy of $\panda$ by 0.4~GeV. 
The cross sections at $\sqrt{s}=5.5$~GeV is predicted as 
$\sigma(\ppb\to\piz\y)=\frac{1}{\Gamma_{\EE}}(29.4\pm0.59\pm5.3)\text{~nb}\cdot\text{keV}$ 
or $\frac{1}{\Gamma_{\EE}}(0.32\pm1.6\pm0.06)\text{~nb}\cdot\text{eV}$, 
where the currently unknown electron partial width of the $\y$ is 
expected in the future.

In summary, we do a combined fit to the $\EE\to \ppbpiz$ cross sections measured by the
BESIII experiment and obtain a better description of the continuum production 
of $\EE\to \ppbpiz$. It shows that the interference effect between continuum and
the resonance amplitudes plays a very important role in determining
the resonance contributions in this mode. BESIII has collected data at much more 
energy points from 4.130 to 4.946~GeV, and the cross sections are expected to be 
further measured and better examined in the future~\cite{white}.


This work is supported in part by National Key Research and
Development Program of China under Contract No.~2020YFA0406301,
Advanced Talents Incubation Program of the Hebei University under 
Contract No. 521100221013, and National Natural Science Foundation 
of China (NSFC) under Contract Nos. 11961141012, 11835012, and 11521505.


\end{document}